\documentclass[11pt]{article}

\usepackage{amsfonts}

\usepackage{amssymb}
\usepackage{latexsym}
\usepackage{graphicx}
\usepackage[english]{babel}
\usepackage[font={small}]{caption}

\usepackage{amsfonts}
\usepackage{latexsym}
\usepackage{graphicx}
\usepackage[english]{babel}
\begin{document}
\input epsf

\def\p{\partial}
\def\h{{1\over 2}}
\def\be{\begin{equation}}
\def\bea{\begin{eqnarray}}
\def\ee{\end{equation}}
\def\eea{\end{eqnarray}}
\def\d{\partial}
\def\la{\lambda}
\def\eps{\epsilon}
\def\bb{\bigskip}
\def\mm{\medskip}
\newcommand{\dm}{\begin{displaymath}}
\newcommand{\edm}{\end{displaymath}}
\renewcommand{\b}{\tilde{B}}
\newcommand{\gm}{\Gamma}
\newcommand{\ac}[2]{\ensuremath{\{ #1, #2 \}}}
\renewcommand{\ell}{l}
\newcommand{\z}{\ell}
\newcommand{\newsection}[1]{\section{#1} \setcounter{equation}{0}}
\def\bb{$\bullet$}
\def\Qbar{{\bar Q}_1}
\def\QPbar{{\bar Q}_p}

\def\q{\quad}

\def\bn{B_\circ}

\let\a=\alpha \let\b=\beta \let\g=\gamma \let\d=\delta \let\e=\epsilon
\let\c=\chi \let\th=\theta  \let\k=\kappa
\let\l=\lambda \let\m=\mu \let\n=\nu \let\x=\xi \let\r=\rho
\let\s=\sigma \let\t=\tau
\let\vp=\varphi \let\vep=\varepsilon
\let\w=\omega      \let\G=\Gamma \let\D=\Delta \let\Th=\Theta
                     \let\P=\Pi \let\S=\Sigma

\def\h{{1\over 2}}
\def\t{\tilde}
\def\r{\rightarrow}
\def\nn{\nonumber\\}
\let\bm=\bibitem
\def\Kt{{\tilde K}}
\def\b{\bigskip}

\let\p=\partial

\begin{flushright}
%OHSTPY-HEP-T-03-012\\
\end{flushright}
\vspace{20mm}
\begin{center}
{\LARGE  Remnants, Fuzzballs or Wormholes?\footnote{Essay awarded an honorable mention in the Gravity Research Foundation essay competition 2014.}}
\\
\vspace{18mm}
{\bf  Samir D. Mathur }\\

\vspace{8mm}
Department of Physics,\\ The Ohio State University,\\ Columbus,
OH 43210, USA\\mathur.16@osu.edu\\
\vspace{4mm}
 March 31, 2014
\end{center}
\vspace{10mm}
\thispagestyle{empty}
\begin{abstract}

The black hole information paradox has caused enormous confusion over four decades. But in recent years, the theorem of quantum strong-subaddditivity has sorted out the possible resolutions into three sharp categories: (A) No new physics at $r\gg l_p$; this necessarily implies remnants/information loss. A realization of remnants is given by a baby Universe attached near $r\sim 0$.  (B) Violation of the `no-hair' theorem by nontrivial effects at the horizon $r\sim M$.  This possibility is realized by fuzzballs in string theory, and gives unitary evaporation. (C) Having the vacuum at the horizon, but requiring that Hawking quanta at $r\sim M^3$ be somehow identified with degrees  of freedom inside the black hole. A model for this  `extreme nonlocality' is realized by conjecturing that wormholes connect the radiation quanta to the hole.

\end{abstract}
\vskip 1.0 true in

\newpage
\setcounter{page}{1}

In 1974 Hawking found that particle pairs are created by the gravitational field  around the horizon  \cite{hawking}. One member of the pair, $b$, escapes to infinity as radiation, while the other member $c$ falls into the hole to reduce its mass. These two particles are in an entangled state
\be
|\psi\rangle_{pair}\sim |0\rangle_b|0\rangle_c+e^{-{E\over T}}|1\rangle_b|1\rangle_c +\dots
\label{one}
\ee
so the entanglement of the  radiation with the remaining hole keeps rising. This leads to a puzzle near the endpoint of evaporation: how can the small residual hole have  the huge degeneracy required to carry this entanglement? 

This was a deep conundrum, but  several doubts about Hawking's derivation  disguised its seriousness. Did Hawking   implicitly assuming an over-simplistic behavior for transplankian modes? Can subleading corrections to Hawking's computation  cumulate over the long evaporation time, and somehow remove the entanglement problem?

\b

{\centerline{\bf The `theorem'}}

\b

It was therefore useful to have Hawking's argument recast as a rigorous `theorem' \cite{cern}. Good slices were used to bypass the transplanckian problem, and the power of strong subadditivity was used to constrain the impact of small corrections.  If 

\b

(i) The evolution of low energy modes at the horizon is within a fraction $\epsilon$ of the semiclassical expectation, and

\b

(ii) Quanta at $r\gg M$ are essentially independent of the hole

\b

then $S_N$, the entanglement entropy at emission step $N$, must keep growing as
\be
S_{N+1}>S_N+\ln 2 -2\epsilon
\label{five}
\ee

\b

This theorem invalidated several proposed resolutions of the puzzle. For example Hawking had conjectured in 2004 (based on an earlier suggestion of Maldacena \cite{eternal})  that subleading saddle points in the path integral would provide small corrections that would solve the problem \cite{hawkingreverse}. But  (\ref{five}) holds for {\it any} source for  corrections, and so this conjecture is false. 

\b

{\centerline{\bf Classifying resolutions of the paradox}}

\b

The theorem (\ref{five}) also  brings additional clarity to the debate by categorizing the  possible resolutions into three types,  based on the {\it length scale} they invoke for nontrivial quantum gravity effects:
\b

(A) We accept conditions (i) and (ii) of the theorem. This is natural if quantum gravity effects are confined to distances
\be
d\lesssim l_p
\label{six}
\ee
Information will be trapped in a remnant, which could (a) explosively evaporate away (information loss) (b) take the form of a baby Universe, or (c) slowly leak away information over times $t\gg M^3$.

\b

\begin{figure}[!]
\begin{center}
\includegraphics[scale=.48]{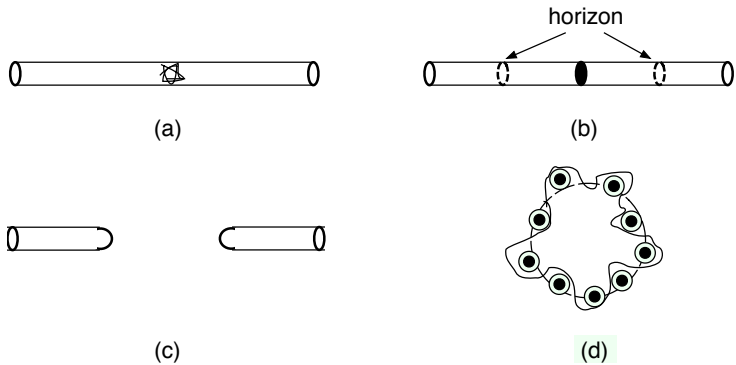}
\caption{{(a) A collection of strings/branes at weak coupling, in a schematic depiction with 1 noncompact and 1 compact dimension. (b) The naive expectation is that at larger coupling a horizon forms. (c) What happens instead is that the compact direction `caps off', so there is no horizon or black hole interior. (d) In 3+1 non-compact dimensions, the analog of this mechanism generates the `fuzzball'. }}
\label{fone}
\end{center}
\end{figure}

(B) Accept (ii) but not (i). This requires us to violate  the `no-hair' theorems for black holes, by finding nontrivial effects at
\be
r\sim 2M
\ee
  `Energy curtains' at the horizon were postulated in \cite{sam}, nonlocality over the horizon scale was suggested in \cite{giddings}, and alterations to the pair production mechanism were invoked in \cite{ellis}. 

But the most explicit mechanism to yield structure at the horizon is found in string theory, where the behavior of extra dimensions  bypasses the assumptions inherent in the no-hair   theorems \cite{gibbonswarner}. Fig.\ref{fone} shows a schematic depiction starting with  one noncompact and one compact space dimension. Fig.\ref{fone}(a)  depicts a large number $N$ of strings and branes when the  coupling constant of the theory is  $g\ll 1$. At large $g$, the naive expectation is that a horizon develops around the central point (fig.\ref{fone}(b)). But in fact the geometry that develops  is the one depicted in fig.\ref{fone}(c): the compact direction pinches off to generate the analogue of a `bubble of nothing' \cite{witten}. The energy of the state is captured by the curvature (and other string fields) at the `caps'. The string theory microstates in 3+1 dimensions are depicted in fig.\ref{fone}(d); the `cap' has the geometry of a Kaluza-Klein monopole, and different orientations of these monopoles at different angular positions reproduce the expected $Exp[S_{bek}]$ microstates of the hole \cite{fuzzballs}.

\b

(C) Accept (i) but not (ii). This requires new physics at length scales
\be
r\gg M
\ee
Papadodimas and Raju \cite{raju} conjecture a long distance non-local interaction, which modifies the state of the radiation quanta $b$ {\it after} they have escaped to distances $r\gg M$ from the hole. Verlinde and Verlinde \cite{ver} have postulated a nonlocal storage of information. 

\begin{figure}[!]
\begin{center}
\includegraphics[scale=.38]{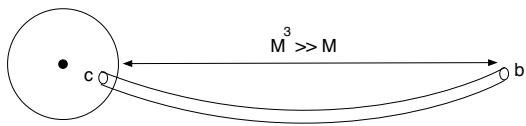}
\caption{{The Maldacena-Susskind conjecture \cite{ms2}: the entangled members of a Hawking pair are connected by a wormhole. }}
\label{ftwo}
\end{center}
\end{figure}

But the most concrete postulate  of this kind was made recently by Maldacena and Susskind \cite{ms2}. They conjectured that whenever two systems are in a quantum entangled state, their locations can be regarded as connected by a spacetime {\it  wormhole}. Since the Hawking radiation quantum $b$ is entangled with its partner $c$ left behind in the hole 
(eq.(\ref{one})), there must be a thin wormhole linking $b$ back to the hole (fig.\ref{ftwo}). Since the radiation extends to distances
\be
 r\sim M^3 \gg M
 \ee
 from the hole during its evaporation, we call this linkage an example of `extreme nonlocality'. With this alteration of spacetime geometry, it is conjectured  in \cite{ms2} that a smooth horizon could be made compatible with unitary Hawking evaporation.
 
 \b

{\centerline{\bf Analysis}}

\b
 
 If we have a fairly minimal theory of quantum gravity, with no extra dimensions or extended objects like strings, then we expect no novel physics for distances $d\gg l_p$. Then we are forced to possibility A.  In this case the Bekenstein entropy would  not be  an  upper bound on the data that can be stored within radius $r=2M$; it would only be   a property of the interface between the hole and its exterior. Hawking has  argued that remnants might also violate CPT \cite{hawking2013}.
  
In string theory, the existence of a dual field theory description forbids information loss.  Further, there appear to be only a finite number of string states at  energy $E\lesssim m_p$; thus there is no place for remnants.  
  
  In 1993, Susskind \cite{susskind1} attempted to avoid remnants, and yet keep {\it both} requirements (a), (b) of the above theorem. His conjecture of `complementarity' argued that the location of information can be {\it observer dependent}. Thus infalling observers see a vacuum at the horizon, while  external observers see information scrambled and reflected back from the horizon. 
  
  Recently, the `AMPS firewall argument' \cite{amps} adapted the bit model of \cite{cern} 
  to prove that Susskind's complementarity was impossible. As a byproduct of this analysis, we can use the  `AMPS gedanken experiment' to distinguish our three possibilities. This experiment  considers an infalling observer who is asked to measure the modes straddling the horizon that would {\it later} evolve to a Hawking pair. The question  is: Are these modes entangled in the manner (\ref{one}), which is appropriate to a local vacuum? For our selected examples from the three categories, we find the answers:
    
  \b
  
  Possibility A (traditional hole): \quad Yes
  
  Possibility B  (fuzzballs): \quad  No
  
  Possibility C (wormholes)\quad  Yes
  
  \b
  
  In particular we see that the fuzzball and wormhole pictures cannot be considered as different descriptions of the same underlying physics.

  \begin{figure}[htbp]
\begin{center}
\includegraphics[scale=.38]{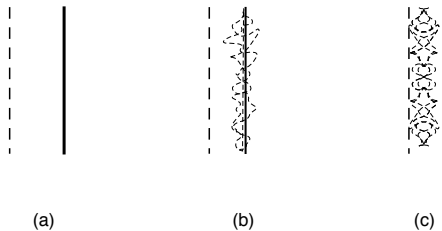}
\caption{{(a) A collapsing shell (solid line) approaches its horizon (dotted line) (b) a part of the wavefunction of the shell spreads over the space of fuzzball states (c) At large Schwarzschild time, the wavefunction has moved out of the shell state and spread completely over the space of fuzzball states.}}
\label{fsixdel}
\end{center}
\end{figure}

\vfill\eject

{\centerline{\bf Fuzzballs vs wormholes}}

\b

There is an enormous body of evidence that the fuzzball mechanism operates for all string states, and we never get a traditional horizon in string theory \cite{fuzzballs}. The wormhole picture, on the other hand,  is suggested by a single example: the eternal hole in AdS, where a wormhole (the Einstein-Rosen bridge) connects two regions of spacetime that are `entangled' \cite{eternal}. It has been recently argued that the dual field theory description of this eternal hole is not as straightforward as had been earlier assumed \cite{difficulties}. In particular, the eternal hole geometry may be unstable to tunneling into disconnected regions, a process that would destroy the Einstein-Rosen bridge \cite{destroy}.

The principal motivation for the wormhole  picture is that it may permit a smooth horizon where the local state is the vacuum. It is natural that one does not want to give up this vestige of our classical intuition, unless there is a good reason to do so.  But the fuzzball picture {\it does} address the relevant questions, as follows:

\b

{\it How does the semiclassical approximation break down at the horizon?}  As a collapsing shell approaches its horizon, there is a small amplitude for it to tunnel into a fuzzball state
\be
{\cal A} \sim e^{-\alpha GM^2}, ~~~\alpha\sim O(1)
\label{el}
\ee
But the number of states to which one can tunnel is large
\be
{\cal N} \sim e^{S_{bek}}\sim e^{4\pi G M^2}
\label{tw}
\ee
The smallness of the tunneling probability $|{\cal A}|^2$ cancels against the largeness of ${\cal N}$. Instead of passing through its horizon, the wavefunction of the shell spreads over the space of fuzzball states (fig.\ref{fsixdel}). 

\b

  \begin{figure}[htbp]
\begin{center}
\includegraphics[scale=.38]{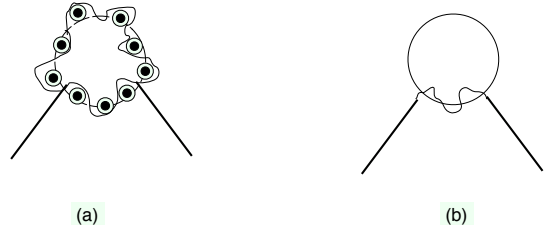}
\caption{{(a) Probing the fuzzball with operators at energy $E\gg kT$ causes collective excitations of the fuzzball surface. (b) The corresponding correlators are reproduced in a thermodynamic approximation by the traditinal black hole geometry, where we have no fuzzball structure but we use the geometry on both sides of the horizon.}}
\label{feetwo}
\end{center}
\end{figure}

{\it Is there any role for the traditional black hole geometry?} The fuzzball has no interior, and the $E\sim T$ quanta emitted from its surface carry information about its state. Under the impact of  $E\gg T$ quanta, however, the fuzzball surface suffers `collective oscillations'. The Green's functions describing these oscillations (fig.\ref{feetwo}(a)) agree to an excellent approximation with  Green's functions measured in the {\it traditonal} black hole geometry (fig.\ref{feetwo}(b)). Since $T\sim M^{-1}$ is very small, typical infalling quanta will feel very little at the horizon, not because they can `pass through', but because the excitations they cause on the fuzzball surface have an approximate  {\it dual} description in terms of the traditional hole \cite{fcomp}.

To summarize, the use of strong subadditivity has brought considerable clarity to the  collection of proposals for resolving the information paradox. In string theory, it appears quite certain that the fuzzball mechanism provides the required resolution, while in traditional quantum  gravity theories (eq.(\ref{six})) remnants are the logical choice.

\end{document}